\documentclass[graybox]{svmult}

\usepackage{mathptmx}       
\usepackage{helvet}         
\usepackage{courier}        
\usepackage{type1cm}        
\usepackage{makeidx}         
\usepackage{graphicx}        
\usepackage{multicol}        
\usepackage[bottom]{footmisc}
\usepackage{twocolumn}
\usepackage{url}

\makeindex             

\begin{document}

\title*{Deep photometry of galaxies in the VEGAS survey: the case of NGC 4472}
\titlerunning{Photometry of NGC4472} 
\author{Marilena Spavone on behalf of the VEGAS team\thanks{
VEGAS team: 
M. Capaccioli (P.I.), Michele
Cantiello, D. A. Forbes, A. Grado, E. Iodice, L. Limatola,
N. Napolitano, M. Paolillo, T. H. Puzia, G. Raimondo, A.
J. Romanowsky, P. Schipani \& M. Spavone}
}
\authorrunning{M. Spavone} 
\institute{INAF-Astronomical Observatory of Capodimonte \at Salita
  Moiariello 16, 80133, Naples, Italy \email{spavone@na.astro.it}}
%
%
\maketitle

\abstract{The VST-VEGAS project is aimed at observing and studying a
  rich sample of nearby early-type galaxies in order to systematically
  characterize their properties over a wide baseline of sizes and out
  to the faint outskirts where data are rather scarce so far. The
  external regions of galaxies more easily retain signatures about the
  formation and evolution mechanisms which shaped them, as their
  relaxation time are longer, and they are more weakly influenced by
  processes such as mergers, secular evolution, central black hole activity,
  and supernova feedback on the ISM, which tend to level age and
  metallicity gradients. The collection of a wide photometric dataset
  of a large number of galaxies in various environmental conditions,
  may help to shed light on these questions. To this end VEGAS
  exploits the potential of the VLT Survey Telescope (VST) which
  provides high quality images of one square degree field of view in
  order to satisfy both the requirement of high resolution data and
  the need of studying nearby, and thus large, objects. We present a
  detailed study of the surface photometry of the elliptical galaxy
  NGC4472 and of smaller ETGs in its field, performed by using new g
  and i bands images to constrain the formation history of this nearby
  giant galaxy, and to investigate the presence of very faint substructures in its surroundings.}

\section{The VEGAS survey}
\label{sec:1}
The VST Elliptical GAlaxies Survey (VEGAS) is a deep multi-band ($g,
r, i$) imaging survey of early-type galaxies in the Southern
hemisphere carried out with VST at the ESO Cerro Paranal Observatory
(Chile). The survey goal is to map the surface brightness of galaxies
with $V_{rad}<4000$ km/s, sampling all environmental conditions and
the whole parameter space. The expected depths at $\mbox{S/N}>3$ in
the $g$, $r$ and $i$ bands are 27.3, 26.8, and 26 mag arcsec$^{-2}$
respectively, enough to detect signatures of diffuse star components
around galaxies (see e.g. \cite{Zibetti05}) and the dynamical
interactions of ETGs with the intergalactic medium. The main aspects
that the VEGAS survey will investigate are: 1) 2D light distribution out to 8-10 $R_e$:
galaxy structural parameters and diffuse light component, inner
substructures as a signature of recent cannibalism events, inner disks
and bars fueling active nuclei present in almost all the objects of
our sample; 2) radially averaged surface brightness profiles and
isophote shapes up to 10 $R_e$; 3) color gradients and the connection
with galaxy formation theories; 4) detection of external low-surface
brightness  structures of the galaxies and the connection with the
environment; 5) census of small stellar systems (SSS: GCs, ultra
compact dwarfs and galaxy satellites) out to $\sim$200 kpc from the
main galaxy center, and their photometric properties (e.g. GC
luminosity function and colors and their radial changes out to several
$R_e$ allowing to study the properties of GCs in the outermost
``fossils'' regions of the host galaxy. VEGAS will provide a volume
limited survey in the South, complementary  to the Next generation Virgo Cluster Survey (NGVS), with comparable depth but no environmental restrictions.

\section{The NGC 4472 field: a test case}
\label{sec:2}
This first VEGAS case (Capaccioli et al., in preparation) deals with a deep photometric analysis of the ETGs in the VST field of the galaxy NGC 4472 (M49),
the brightest member of the Virgo cluster.
This field has been chosen for the following reasons:
\begin{itemize}
\item it is well-studied, with an ample scientific photometric literature (\cite{Ferrarese06}, \cite{Mihos13}, \cite{Kormendy09}, \cite{Janowiecki10}, \cite{Kim00});
\item it offers a wide range of cases where to investigate the ability of VEGAS to map the faint galaxy outskirts.
In fact, together with a nearby supergiant object filling almost the entire OmegaCAM field, there are smaller ETGs either embedded in the light of NGC 4472 or close to the edges of the frame.
\end{itemize}

\subsection{Light and color distribution}
\label{subsec:1}
\begin{figure}[b]
\vspace{-25 pt}
\sidecaption
\includegraphics[width=7cm]{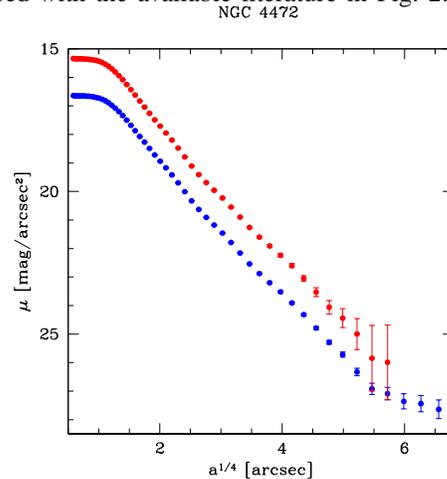}
\vspace{-37 pt}
\caption{Azimuthally averaged light profiles for NGC 4472 in the {\it g} (blue) and {\it i} (red) bands.}
\label{prof}       
\end{figure}

We used the standard {\small IRAF}\footnote{IRAF is distributed by the
  National Optical Astronomy Observatories, which is operated by the
  Associated Universities for Research in Astronomy, Inc. under
  cooperative agreement with the National Science Foundation.({\it
    Image Reduction and Analysis Facility}) environment.} task {\small
  ELLIPSE} to perform the isophotal analysis of the VEGAS galaxies on
the final mosaic in each band, after proper background subtraction. We
have also derived the azimuthally averaged surface brightness profiles
in isophotal annuli of specified thickness. The azimuthally averaged
profiles for NGC 4472 in the {\it g} and {\it i} bands are shown in
Fig.\ref{prof} as a function of the isophote semi major axis $a$.

The light profile in the {\it g} band, shows a neat
change in the slope at $a_{e}^{1/4} \simeq\ 5.5$, where $\mu_{g} \sim
27$ mag arcsec$^{-2}$. The level at which the break occurs is compatible with the typical values at which \cite{Zibetti05}
have observed a change of slope induced by the ICL in a series of stacked 
galaxy clusters.  Moreover we note the presence of an outer and
more elliptical component with a significant gradient in the P.A.
which, as suggested by \cite{Gonzalez05}, is likely due to a
population of some ICL.

\begin{figure}[b]
\vspace{-30 pt}
\sidecaption
\includegraphics[width=6cm]{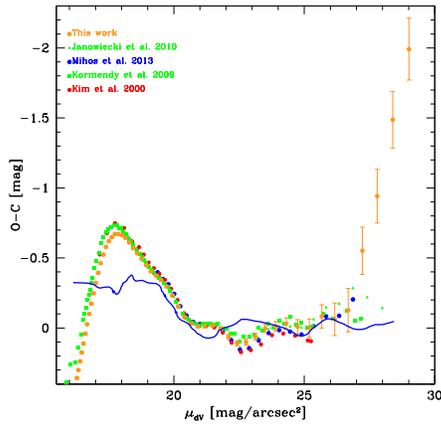}
\vspace{-10 pt}
\caption{(O-C) residuals of mean profiles from
     a best fitting $r^{1/4}$ model used only to remove the main
     gradient and make comparisons easier. The blue solid
     line plots the O-C residuals for the East-West photometric cross-section of
     the standard elliptical galaxy NGC3379 from \cite{deV79}.}
\label{conf}       
\end{figure}

Our azimuthally averaged {\it g} band profile is compared with the
available literature in Fig. \ref{conf}. The residuals with respect to $r^{1/4}$ fits show a
spectacular agreement with the literature, and from the comparison with
NGC3379 (\cite{deV79}, blue line) we note clear similarities due to
the presence of diffuse shells in both systems. 

We have also studied the fainter ETGs in the one square degree of the
OmegaCAM field: NGC 4434, NGC 4464, NGC 4467, and VCC 1199, including the dwarf irregular, UGC 7636 in the proximity of the giant NGC 4472, reaching an even larger depth for these systems.

\begin{figure}[b]
\vspace{-30 pt}
\sidecaption
\includegraphics[width=6cm]{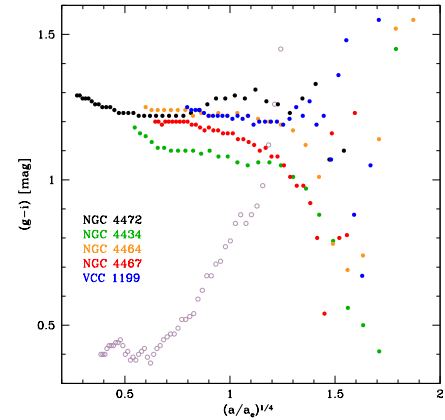}
\vspace{-10 pt}
\caption{\emph{Left}-Azimuthally averaged light profiles in the {\it g} band for the five ETGs of this paper scaled to their effective parameters. \emph{Right}-Assembly of the {\it (g-i)} color profiles for the five ETGs and for the interacting system UGC 7636.}
\label{all}       
\end{figure}

The (g-i) color profiles show an indication that for $r >3
r_{e}$ a very negative colour gradient develops in some galaxies,
which apparently vanishes at $r \simeq\ 8 r_{e}$ (see Fig.\ref{all}).

\subsection{2-dimensional model}
\label{subsec:2}
\begin{figure}[b]
\vspace{-20 pt}
\sidecaption
\includegraphics[width=6cm]{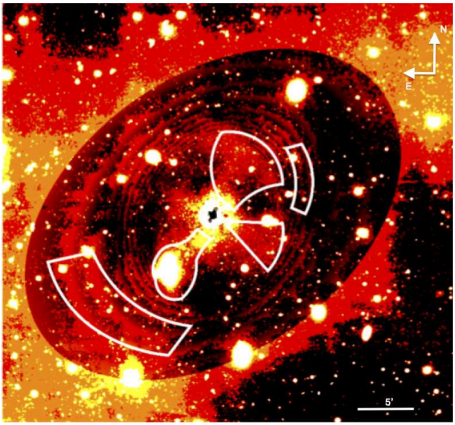}
\caption{NGC 4472. Zoom ($37 \times\ 35$ arcmin) of the median-smoothed residual image. Superimposed white contours are 1) the tail
    connecting UGC 7636 to the giant ETG, and 2) shells and fan of
    material identified by \cite{Battaia12} and visible in our
    residual image.}
\label{shell}       
\end{figure}

In order to enlighten possible larger substructures, a 2-dimensional elliptical
model of NGC4472 best fitting the azimuthally averaged isophotes has
been produced using the {\small IRAF} task {\small BMODEL}.
Fig. \ref{shell} plot the difference between the original {\it g} band
image and the model.
This residual map shows a clear asymmetry in the nuclear region and some diffuse features such as a tail
associated with the dwarf irregular galaxy UGC 7636 interacting with NGC 4472, as well as the presence of concentric shells
and fans of material  (white contours), also identified by
\cite{Battaia12}.

\section{Conclusions}
\label{sec:3}
We have presented the VST Early-type Galaxy Survey (VEGAS), currently
ongoing with VST/OmegaCAM (PI: M. Capaccioli). In
particular, we present the deep observations in two bands ({\it g} and
{\it i}), collected with the VST/OmegaCAM, for NGC4472. The surface brightness profiles of NGC 4472 reach a depth of $27.5$ mag/arcsec$^2$ in {\it g} band and $26$ mag/arcsec$^2$ in {\it i} band, comparably to previous deep studies (see Fig. \ref{conf}).
This depth allowed us to spot deviations from a simple de Vaucouleurs profile and in particular a change of slope at $a \sim14'.2$ (see Fig. \ref{prof})
that we have associated to the presence of a decoupled ICL component which was not detected
in previous analyses.

Here we stress that the simple inspection
of the deep surface brightness profile of NGC 4472, clearly shows the presence of a diffuse component starting to dominate 
at $\mu_{g} \sim$26.5 mag/arcsec$^2$ (see Fig. \ref{prof}), which is compatible with the typical values at which \cite{Zibetti05}
have observed change of slope induced by the ICL in a series of
stacked galaxy clusters.

We notice that the trend of the residuals of the luminosity profiles
of NGC 4472 with respect to an $r^{1/4}$ best fitting model has some
striking analogies with the similar curve for NGC 3379
(\cite{deV79}). Besides a bright extended core, we find evidence for a
wavy pattern possibly associated with shells of diffuse material.
The presence of such shells also stand out very clear from the 2-D residual map. 

We have also studied the fainter ETGs in the one square degree of the
OmegaCAM field: NGC 4434, NGC 4464, NGC 4467, and VCC 1199, including the dwarf irregular, UGC 7636 in the proximity of the giant galaxy NGC 4472. For all
these systems we have highlighted the presence of some substructures
defined as deviations from a simple de Vaucouleurs (1948) best fit
profile. These deviations are associated to strong varying values of
the ellipticity and P.A. as well as $a_4$ and $b_4$ parameters,
suggesting the presence of some substructures.

The color profiles show an indication that for $r >3 r_{e}$ a very
negative colour gradient develops in some galaxies, which apparently
vanishes at $r \simeq\ 8 r_{e}$.
In a forthcoming paper (Capaccioli et al., in preparation) we will
show that our results are basically unaffected in their qualitatively
conclusions by the extended wings of the PSF (scattered light).

To conclude, this paper illustrates the performance and the accuracy
achieved with the VST/OmegaCAM to produce surface photometry of
early-type galaxies, also in very extreme conditions. For the case of
NGC 4472 the presence of an extended halo around the giant galaxy, reaching the edge of the one square degree field of view, has allowed us to fully test the procedure for data reduction and background subtraction.  
The results obtained with our observations are comparable for accuracy
to the collection of observations gathered from different telescopes
(see \cite{Kormendy09}). 

In the future we expect to implement a more
variegate surface analysis including a wider set of photometric law in
order to characterize the SB measurements in a larger sample of
galaxies and thus discuss results in the context of galaxy formation
theories.
\vspace{-20 pt}

\end{document}